\documentclass{article}
\usepackage[T1]{fontenc}

\usepackage[default,regular,black]{sourceserifpro}
\usepackage{sectsty}
\usepackage{siunitx}
\usepackage[english]{babel}

\usepackage[letterpaper,top=2cm,bottom=2cm,left=3cm,right=3cm,marginparwidth=1.75cm]{geometry}

\usepackage{amsmath}
\usepackage{graphicx}
\usepackage[colorlinks=true, allcolors=blue]{hyperref}
\title{
On the addition of micron-size intruders in a shear-thickening suspension of nanoparticles}
\author{
Alice Pelosse \& Heinrich M. Jaeger
\vspace{0.2cm}\\
\textit{\small James Franck Institute, University of Chicago, Chicago, Illinois 60637, USA}}
\usepackage [T1] { fontenc } 
\usepackage { tgbonum }
\usepackage{natbib}
\begin{document}
\maketitle

\begin{abstract}
This study investigates the rheological behavior of shear-thickening suspensions made of different types of nanoparticles upon the addition of large intruders referred to as granules. 
The size ratio ranges from 20 to 120.
We examine the effects of granule size, volume fraction, and surface properties on shear-thickening characteristics. 
Starting with a fumed silica suspension exhibiting discontinuous shear thickening (DST) without granules, the addition of granules at different volume fractions, shifts the onset of thickening to lower shear rates.
Concomitantly, the strength of the thickening, quantified by the thickening index, decreases, transitioning from DST to continuous shear thickening (CST).
Comparison with suspensions of nanosilica spheres reveals a similar trend, suggesting generality across different systems. 
However, these results contrast with cornstarch-based suspensions, where granule addition enhances thickening. 
This difference is attributed to the large size ratio studied here: When the granules are much larger than the particles in the interstitial suspension, the granules introduce a spread in the local shear rate and disrupt the particles' ability to form an extended fabric of force chains. 
The findings highlight the critical role of particle size ratio in determining the rheology of complex suspensions, paving the way for tailoring material properties in industrial and scientific applications.
\end{abstract}
\section{Introduction}

Shear-thickening of a suspension, i.e., the increase of viscosity with strain rate or applied stress, is observed in a plethora of systems, such as starches, dense colloidal and confined granular suspensions, or suspensions of functionalized particles.
This phenomenon usually appears at relatively high particle volume fractions and is thus intrinsically related to the jamming of the solid phase \citep{mari2014shear, morris2020shear}.

Several mechanisms can explain such a rheology and contribute in different proportions to an observed shear-thickening behavior \citep{brown2014shear}, which can result in continuous shear thinning (CST) or discontinuous shear-thickening (DST).
The transition from an ordered to a disordered state has been evidenced for highly monodisperse particles while increasing the flow shear rate and is concomitant with an increase in the viscosity \citep{hoffman1974discontinuous, lee2018unraveling}.
This increase can be smooth or very abrupt, but this mechanism cannot be sufficient to account for shear-thickening in most suspensions, where ordering and crystallization do not occur due to particle size variation.
Another mechanism leading to shear-thickening is that of \textit{hydroclusters}, coming from hydrodynamic interactions between the particles through viscous pressure as they are brought close.
These hydroclusters are therefore transient aggregates of particles, formed as a sufficiently large shear rate is applied on the system \citep{brady1985rheology}.
They grow in size with increasing volume fraction and shear rate (or Peclet number), correlated with the shear-thickening in the suspension \citep{cheng2011imaging}.
However, such mechanism seems limited to CST and cannot account for DST \citep{brown2014shear}.
Instead of hydroclusters, physical and chemical friction between the particles in concentrated suspension is able to create force chains which percolate in the system, resist shear, and can lead to DST rheology of the suspension \citep{seto2013discontinuous}.
In that case, additional complexity arises from the solvent, which mediates hydrodynamic interactions but also specific surface interactions of the particle surfaces \citep{van2021role}.
The frictional contribution to dissipation relative to hydrodynamic lubrication depends on the applied shear stress and inter-particle friction and controls the CST/DST behavior of the suspension \citep{fernandez2013microscopic, royer2016rheological,more2020roughness}.
The activation of frictional contacts is correlated with the onset of thickening for increasing shear rate or stress \citep{wyart2014discontinuous}.
This creates force chains and whole clusters of frictionally coupled particles, whose size and velocity correlation grow with applied shear \citep{seto2013discontinuous,heussinger2013shear}. 
This increased dissipation mechanism depends on the friction strength  \citep{clavaud2017revealing} but also on the confinement stress that will trigger DST when a system is frustrated and cannot dilate to rearrange \citep{brown2012role}.
In addition to increasing friction, particle roughness and asperities can yield particle interlocking, thereby further decreasing both the shear rate and the particle volume fraction required to turn on shear-thickening \citep{hsu2018roughness,blair2022shear}.

Non-Brownian cornstarch suspensions have been extensively studied and exhibit strong DST at relatively low volume fraction.
Brownian suspensions of nanoparticles can be shear-thickening depending on the nature and strength of the frictional effects aforementioned.
In particular, fumed silica particles, which are aggregates of silica nanospheres and have fractal-like shapes of size ranging from 100 to \SI{500}{\nano\meter}, can exhibit strong DST \citep{raghavan1997shear,bourrianne2022tuning}.
Suspensions of non-aggregated silica nanospheres in appropriate solvents can also exhibit CST/DST behavior \citep{cwalina2014material}.
Simulations suggest that thermal effects would mainly shift the shear-thickening onset to higher values, with Brownian forces adding to repulsive interactions between the particles \citep{mari2015discontinuous, kawasaki2018discontinuous}.

In this article, we investigate how the addition of large intruders with no specific ability to trigger shear-thickening, here termed \textit{granules}, disturb the shear-thickening of a colloidal suspension used as a \textit{solvent}.
The addition of these granules to a non-Newtonian solvent is relevant for many industrial processes (concrete) and natural systems (blood) and raises non-trivial questions. 
In the low-inertia regime, the viscosity of a suspension made solely with these granules is relatively simple when the suspending fluid is Newtonian.
In that case, the viscosity of the suspension is proportional to the suspending fluid viscosity, and the relative viscosity of the  suspension to the fluid solvent, $\eta_r$, is solely a function of the granule volume fraction, $\phi_\text{G}$ normalized by the jamming fraction $\phi_\text{G}^c$, which depends on the particle properties \citep{guazzelli2018rheology}.
In particular, the relative viscosity of granular suspensions with different polydispersity \citep{chong1971rheology, cheng2011imaging, pednekar2018bidisperse} and surface roughness \citep{tapia2019influence} can be collapsed onto the same master curve which diverges as jamming is approached, i.e. as $\phi_\text{G}/\phi_\text{G}^c\rightarrow 1$.
This  geometrical effect of polydispersity saturates for size ratio larger than 10, due to the saturation of the jamming fraction when large particle packing is no more affected by addition of small particles \citep{chong1971rheology}.

For large size ratio, e.g. granules in a colloidal suspension, the granule environment can then be approximated by a continuous phase with a non-Newtonian viscosity \citep{sengun1989bimodal}.
For that reason, adding granules in a non-Newtonian fluid should increase the zero shear viscosity by a factor close to the relative viscosity, $\eta_r$.
Adding granules to this solvent raises questions about the definition of the shear rate, which is locally enhanced upon addition of granules.
This enhanced shear rate is captured by a lever function and depends on the granule volume fraction only \citep{dagois2015rheology,guazzelli2018rheology}.

Experiments where larger granules were added to a non-Brownian, shear-thickening cornstarch suspension have reported both effects, namely (i) the increase in the zero-shear viscosity and (ii) the shift of the onset of shear thickening to smaller shear rate \citep{madraki2017enhancing}.
Importantly, in these systems a transition from  CST to  DST behavior was observed. 
This was explained as a wall-effect whereby volume exclusion near the surface of the large granules locally increased the volume fraction of the surrounding shear thickening suspension, resulting in  increased dissipation of the interstitial fluid \citep{madraki2018transition}.
The latter is thus a finite size effect arising when granule  size is larger, but still close to that of the particles in the suspending fluid.
In this article, we discuss the opposite limit, namely a large size ratio, ranging from 20 to 120, between micron-size granules and nanoparticles in the solvent.
We show that in this limit, we reach similar results for the critical shear rate and shear stress to obtain shear-thickening, but opposite conclusions for the strength of the shear-thickening upon addition of granules. 
More specifically, instead of enhancing shear thickening we find a transition from DST to weaker CST behavior when adding granules.
Interestingly, the granule size still matters despite the large size contrast in our systems.

The article is organized as follows.
In the next section (\S~\ref{sec:methods}), we present the different suspensions and protocols we used in experiments.
The results are described in (\S~\ref{sec:results}) for the different suspensions.
A discussion and comparison with other systems found in literature is then presented in (\S~\ref{sec:discussion}) and a model to account for our observations is drawn.

\section{Material and Methods}
\label{sec:methods}
\textbf{Shear thickening suspensions: }Hydrophilic fumed silica particles (AEROSIL OX50, $\rho_\text{p}=$~\SI{2.2}{\gram\per\cubic\centi\meter}) are dispersed in liquid PEG ($M_\text{w} = $~\SI{200}{\gram\per\mole}, $\rho_\text{f}=$~\SI{1.12}{\gram\per\cubic\centi\meter}, $\eta_\text{f}=$~\SI{57}{\milli\pascal\second} measured at \SI{22}{\celsius}) purchased at Sigma Aldrich.
To ensure homogeneous dispersion, the fluid was weighed, and the desired amount of particles was added progressively in three steps. 
After each addition, the mixture was subjected to sonication, followed by manual and mechanical mixing. 
The final suspension was further sonicated and mixed using a rolling device for one week, ensuring it reached a stable state.

Spherical colloidal silica suspensions were prepared following the same protocol, using particles with a diameter of \SI{500}{\nano\meter} (AngstromSphere, $\rho_\text{p} = $~\SI{1.8}{\gram\per\cubic\centi\meter}) in the same PEG solvent. 
Similarly, an extended mixing period of one week was necessary to achieve a uniform dispersion of nanoparticles.

The nanoparticle volume fraction of the shear-thickening suspension is defined as the concentration excluding intruding particles, $\phi_\text{STF} = V_\text{NP}/(V_\text{NP}+V_\text{f})$, where $V_\text{NP}$ and $V_\text{f}$ denote the volumes of nanoparticles and fluid, respectively. 
Note that, under this definition, $\phi_\text{STF}$ represents the concentration within the interstitial fluid and not in the total suspension after the introduction of larger intruding particles.
The nanoparticle volume fraction and proportion in the solid phase thus decreases upon addition of granules, see an example for $\phi_\text{STF} = 30~\%$ in Table~\ref{tab1}.

\begin{table}
  \begin{center}
    \label{tab:table1}
    \begin{tabular}{lccccc}
    \hline
      \\[-1em]
      Granule volume fraction $\phi_\text{G}$ (\%) &0 & 20  &30 &40 &45\\
      \\[-1em]
      \hline \hline
      \\[-1em]
      Nanoparticle volume fraction in total suspension $\phi_\text{NP}$ (\%) & 30 & 24& 21 & 18 &17\\
      \\[-1em]
      Total solid volume fraction (\%) & 30 & 44 & 51 & 58 & 62\\
      \\[-1em]
      Nanoparticle proportion in solid volume $\alpha$ (\%) & 100 & 54 & 41 & 31 & 27\\
      \\[-1em]
      \hline
    \end{tabular}
  \end{center}
      \caption{Different volume fractions of a system in which granules are added to a nanoparticle suspension with $\phi_\text{STF} = 30~\%$.}
    \label{tab1}

\end{table}

\textbf{Granules: }
Large particles, referred to as intruders or granules, were selected to test various properties, including size, density, and surface characteristics.
The granules used included hollow glass spheres with significant polydispersity (Supelco, Sigma-Aldrich, $\rho_\text{G}=$~\SI{1.0}{\gram\per\cubic\centi\meter}, 9-\SI{13}{\micro\meter}), highly monodisperse PMMA spheres (CA20 and CA60, Microbeads, $\rho_\text{G}=$~\SI{1.2}{\gram\per\cubic\centi\meter}, \SIlist{20;60}{\micro\meter}) and highly monodisperse polystyrene spheres (TS20, Microbeads, $\rho_\text{G}=$~\SI{1.1}{\gram\per\cubic\centi\meter}, \SI{20}{\micro\meter}).  
These granules were added to the shear-thickening suspension at volume fractions ranging from 20\% to 45\%. 
The granule volume fraction, $\phi_\text{G}$, is defined as the ratio of the granule volume to the total suspension volume, $\phi_\text{G} = V_\text{G}/(V_\text{G}+V_\text{STF})$, where $V_\text{G}$ is the granule volume, and $V_\text{STF}$ is the volume of the shear-thickening suspension.

\textbf{Rheology protocol: }
Rheological measurements were conducted using a stress-controlled rheometer (Anton Paar, MCR301) with a 25-\si{\milli\meter} plate-plate geometry, maintained at a controlled temperature of \SI{22}{\celsius}.
To prevent slippage when measuring samples containing the largest granules, sandpaper was taped on the plates.
The gap height was set to be at least 10 times larger than the size of the largest particles, typically ranging from \num{0.5} to \SI{1}{\milli\meter}.
Prior to measurements, suspensions were pre-sheared at a stress of \SI{1}{Pa} for \SI{60}{\second}. Subsequently, they were subjected to an upward stress ramp, typically ranging from \SI{1}{\pascal} to 1000-\SI{10000}{\pascal}.
With the high fractions in granules, experiments were stopped at lower shear stress to avoid suspension loss outside of the plate.
Good reproducibility was observed, with consistent results across different samples (typically three).

\section{Results}
\label{sec:results}
\begin{figure}
    \centering
    \includegraphics[width=\linewidth]{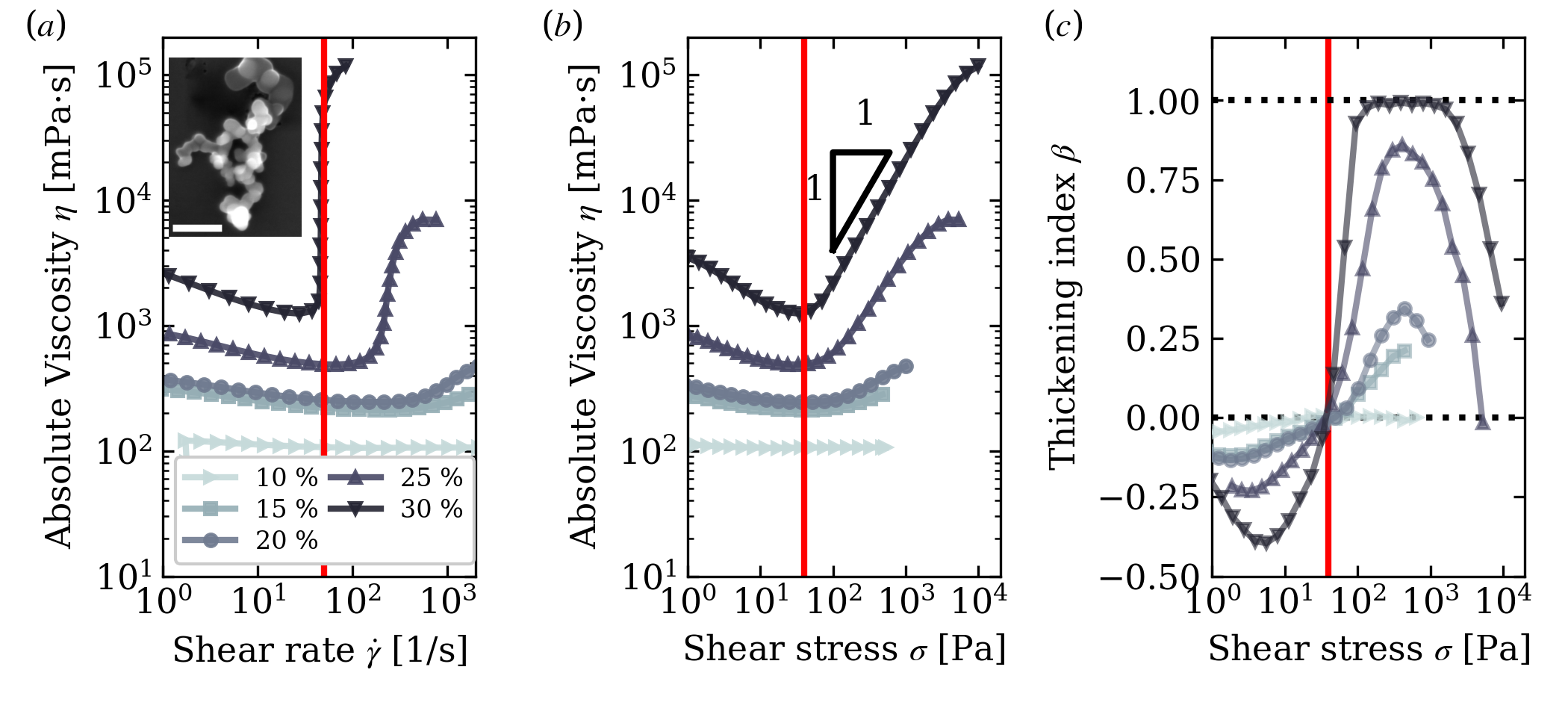}
    \caption{Steady shear rheology of fumed silica suspensions in PEG200 for various volume fractions, ranging from $\phi_\text{STF} = 10~\%$ to $30~\%$.
($a$) Viscosity as a function of shear rate.
($b$) Viscosity as a function of shear stress.
($c$) Thickening index as a function of shear stress.
Red vertical lines indicate the onset of discontinuous shear thickening (DST) for the most concentrated suspension, with $\phi_\text{STF} = 30~\%$.}
    \label{fig1}
\end{figure}

\textbf{Shear-thickening fluid without granules -}
In Fig.\ref{fig1}($a,b$), the steady shear viscosity of a fumed silica suspension is plotted as a function of the shear rate and shear stress for different volume fractions, ranging from $\phi_\text{STF}=$ 10~\% to $\phi_\text{STF}=$~30~\%.
For the highest volume fraction, $\phi_\text{STF} = 30~\%$, the suspension exhibits a discontinuous increase in viscosity at a critical shear rate $\dot{\gamma}_\text{c,0} =$~\SI{50}{\per\second} and at a critical stress $\sigma_\text{c} =$~\SI{40}{\pascal}.
This discontinuous behavior is also captured by the log-log slope of the viscosity versus shear stress, defined as $\beta = d \ln(\eta)/d \ln(\sigma) = \dot{\gamma}(d\eta/d\sigma)$ (see Fig.~\ref{fig1}($c$)).
This parameter, referred to as the thickening index, is negative in the shear-thinning regime, zero for Newtonian fluids, and positive for shear-thickening fluids. 
In the case of discontinuous shear thickening (DST), $\beta \rightarrow 1$.

For all volume fractions, $\beta$ is negative at low shear stress, corresponding to the shear-thinning behavior observed in Fig.\ref{fig1}($a$).
The thickening index then becomes positive as shear stress/rate are increased. 
For the highest concentration, $\phi_\text{STF}=$~30\%, $\beta$ changes sign at $\sigma_\text{c} =$~\SI{40}{\pascal} and plateaus at $\beta =1$ over a large range of shear stress, see Fig.\ref{fig1}($b,c$).
At lower fumed silica concentrations, the shear thickening is less pronounced, as evidenced by the thickening index, which does not reach unity (Fig.~\ref{fig1}($b,c$)).
For these suspensions, a point of maximum thickening can be identified, characterized by a shear rate $\dot{\gamma}_\text{max}$, a shear stress $\sigma_\text{max}$, and a thickening index $\beta_\text{max}$.
In the case of DST, $\dot{\gamma}_\text{max} = \dot{\gamma}_\text{c}$, $\sigma_\text{max} = \sigma_\text{c}$, and $\beta_\text{max} = 1$.\\

\begin{figure}
    \centering
    \includegraphics[width=0.5\linewidth]{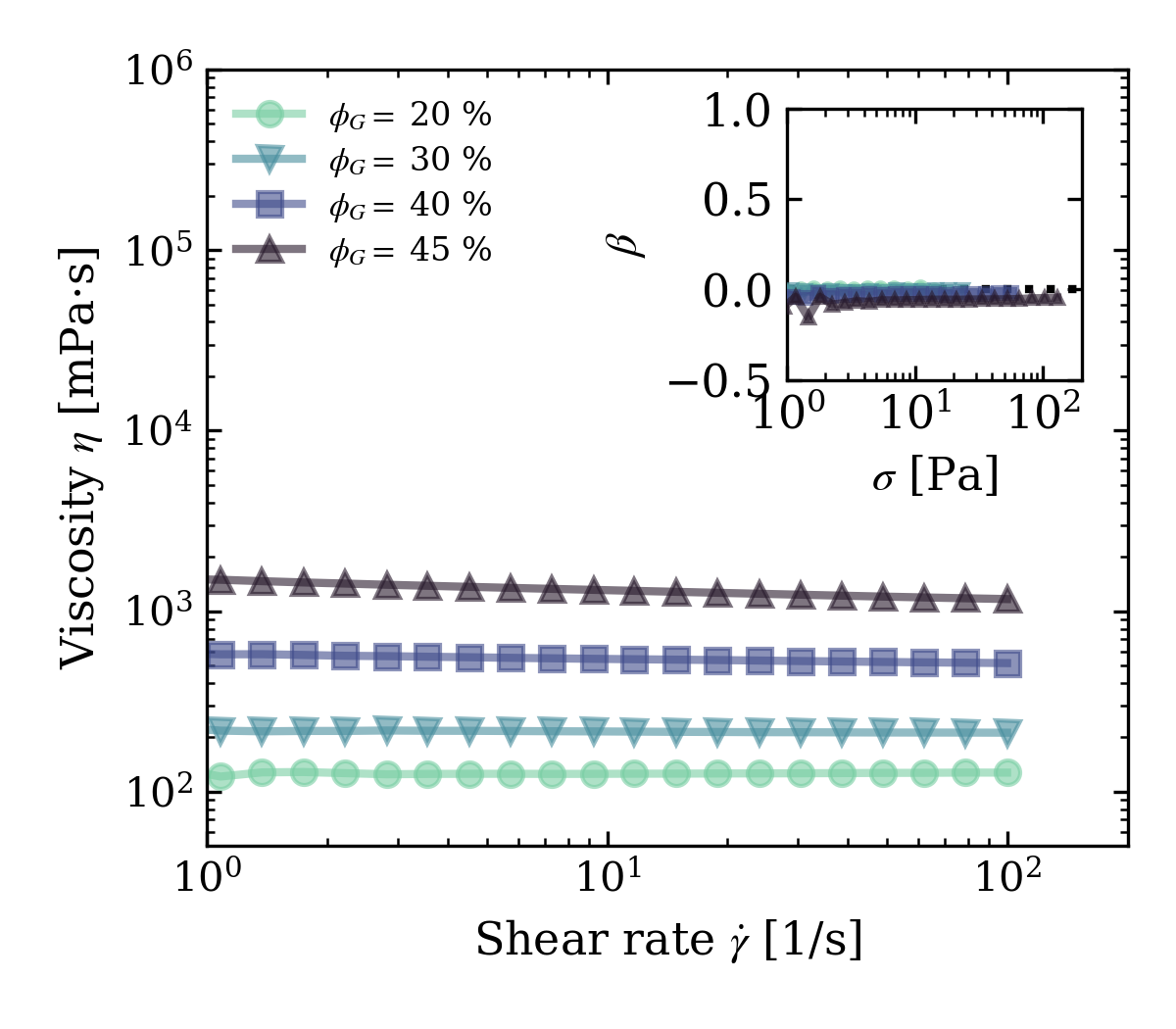}
    \caption{Steady shear rheology of suspension made with 20-\si{\micro\meter} PS spheres in PEG200, for granule volume fractions ranging from $\phi_\text{G} =20~\%$ to $\phi_\text{G} =45~\%$. Main graph: viscosity as a function of shear rate. Inset: thickening index as a function of the shear stress. Relative viscosity $\eta_\text{r} = \eta/\eta_\text{f}$ for increasing granule volume fraction: 2.2, 3.8, 9.6, 23. }
    \label{fig2}
\end{figure}

\textbf{DST fluid and 20-um granules -}
We now turn to mixtures of shear-thickening suspensions with large "granule" intruders, with variable granule size, concentration, and surface properties. 
We remind the reader that these particles in a Newtonian solvent do not shear-thicken, and exhibit a Newtonian behavior and/or a shear-thinning behavior at high shear rate, as expected for such suspensions of non-Brownian particles, see Fig.\ref{fig2}.

\begin{figure}
    \centering
    \includegraphics[width=\linewidth]{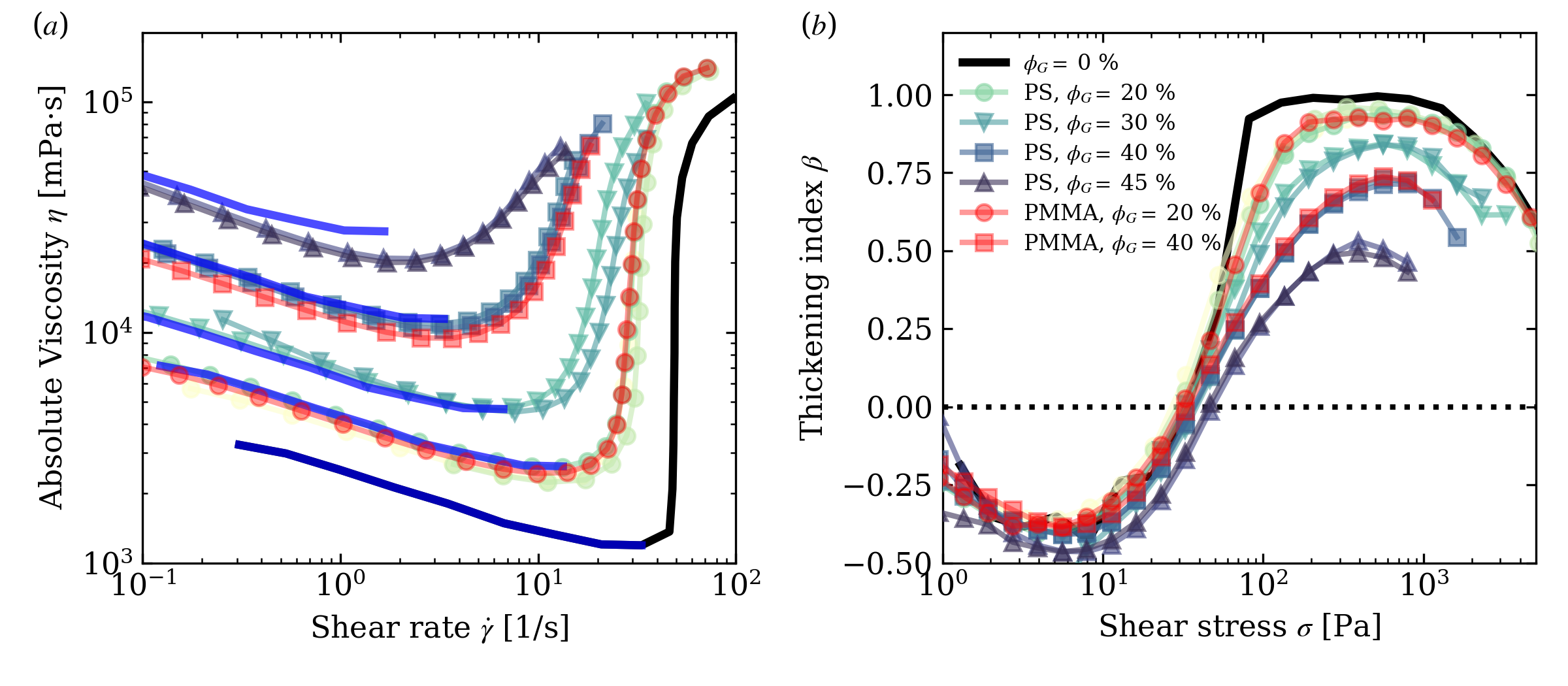}
    \caption{Steady shear rheology of suspensions composed of a fumed silica suspension in PEG200 ($\phi_\text{STF} = 30~\%$) and 20-\si{\micro\meter} polystyrene (PS) spheres, for granule volume fractions ranging from $\phi_\text{G} = 20~\%$ to $\phi_\text{G} = 45~\%$.
($a$) Viscosity as a function of shear rate.
($b$) Thickening index as a function of shear stress.
Blue lines represent the low-shear viscosity prediction based on the relative viscosity of granular suspensions, $\eta_r$ measured in Fig.\ref{fig2}, with the shear rate shifted to align with the onset of shear thickening.
Red curves correspond to suspensions prepared using \num{20}-\si{\micro\meter} PMMA spheres instead of PS.
Black curves represent the baseline behavior of the pure fumed silica suspension extracted from Fig.\ref{fig1} at $\phi_\text{STF} = 30\,\%$.}
    \label{fig3}
\end{figure} 

In Fig.~\ref{fig3}($a$), the viscosity of suspensions containing varying amounts of 20-\si{\micro\meter} PS and PMMA spheres is plotted as a function of the shear rate.
The suspending fluid is a shear-thickening suspension of fumed silica at a volume fraction $\phi_\text{STF} = 30,\%$.
At this particle size, the density of the granules does not significantly affect the suspension behavior, as evidenced by the good overlap of the PS and PMMA rheology data for granule volume fractions $\phi_\text{G} = 20\,\%$ and $40\,\%$.

As the granule fraction increases from $\phi_\text{G} = 20\,\%$ to $\phi_\text{G} = 45\,\%$, two trends are observed in Fig.~\ref{fig3}($a$):
(i) The low-shear viscosity increases, which is expected due to the additional granule loading. 
This behavior is consistent with the scaling of non-Brownian suspensions, where the viscosity is the product of the suspending fluid viscosity and the suspension’s relative viscosity, $\eta_r$ that can be extracted from Fig.\ref{fig2}.
(ii) The shear rate corresponding to the onset of shear thickening decreases with increasing $\phi_\text{G}$, shifting the viscosity curves toward lower shear rates. 
This shift arises from local effects: the granules amplify the local shear rate in the interstitial fluid, triggering shear thickening at smaller applied shear rates, $\dot{\gamma}_\text{c}(d_\text{G}, \phi_\text{G})$.
Shifting the low-shear suspending fluid viscosity by $\dot{\gamma}_\text{c}(d, \phi_\text{G})$ along the $x$-axis and multiplying it by the relative viscosity $\eta_\text{r}(\phi_\text{G})$ along the $y$-axis shows a good collapse in the shear thinning regime, see the blue curves in Fig.~\ref{fig3}($a$).

A third observation is the softening of the shear thickening as the granule loading increases. 
The latter point is not trivial and in contradiction with previous results obtained in a different shear-thickening fluid. 
To best quantify the softening in thickening, the thickening index and the different suspensions is plotted in Fig.~\ref{fig3}($b$).
While the suspensions exhibit similar rheological behavior in the shear-thinning regime ($\beta < 0$), the thickening index decreases with increasing granule content in the shear-thickening regime ($\beta > 0$). The effect of granules on $\beta$ seems to appear right after the shear-thickening onset. 
At $\phi_\text{G} = 20,\%$, the suspension nearly exhibits discontinuous shear thickening (DST) with $\beta_\text{max} \approx 1$. 
The thickening is however mildly weakened considering that a higher shear stress is required to reach this value.
At higher granule fractions, the thickening transitions to a continuous shear-thickening (CST) regime with $\beta_\text{max} < 1$.

Finally, the shear stress at the onset thickening, $\sigma_\text{c}\simeq 40$~\si{\pascal}, measured for $\beta = 0$, seems to be fixed at the critical shear stress of the pure fumed silica suspension (within experimental uncertainty), see Fig.~\ref{fig1}($b$) and Fig.~\ref{fig3}($b$), contrary to the critical shear rate which decreases with increasing $\phi_\text{G}$, see Fig.~\ref{fig3}($a$).\\

\textbf{DST fluid and other granules -}
\begin{figure}
    \centering
    \includegraphics[width=\linewidth]{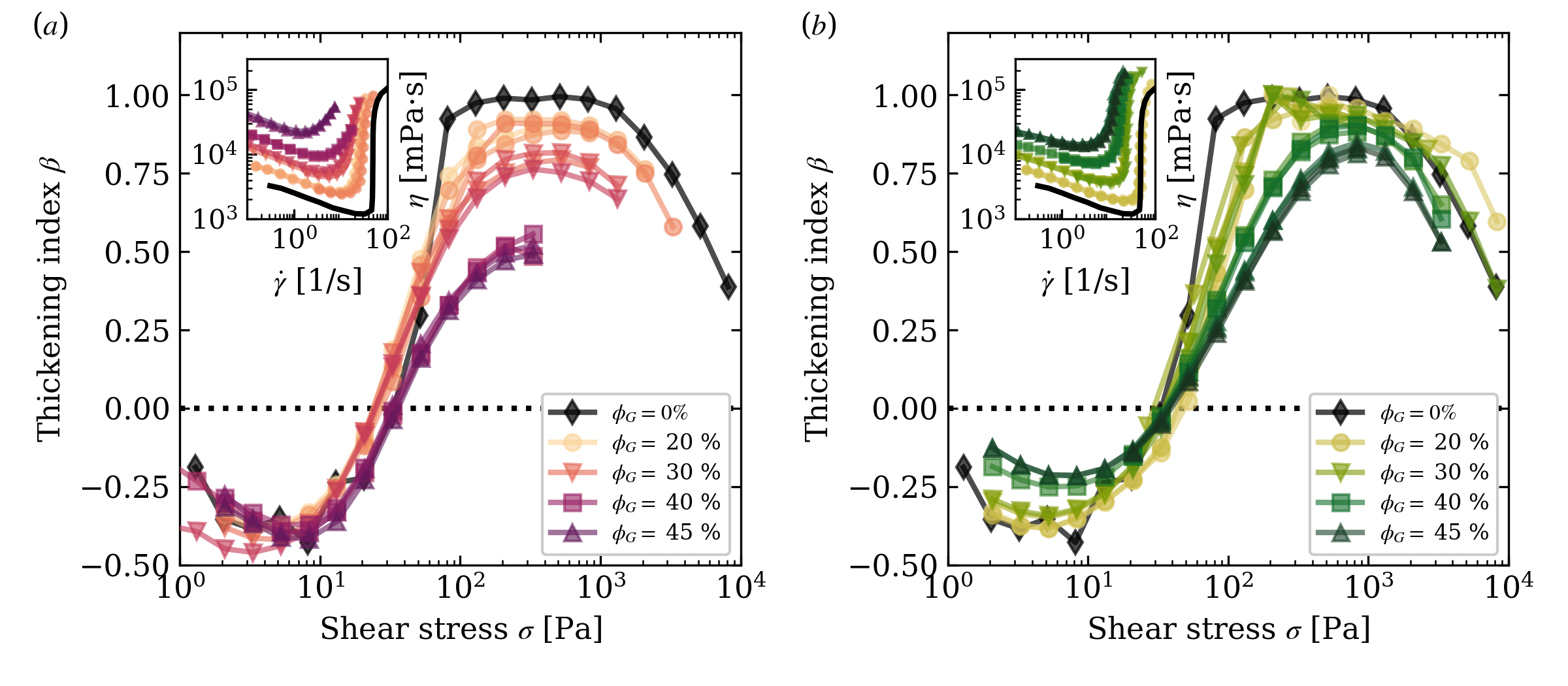}
    \caption{Steady shear rheology of suspension made with a fumed silica suspension in PEG200, $\phi_\text{STF} =30~\%$ combined with ($a$) 10-\si{\micro\meter} glass spheres, and ($b$) 60-\si{\micro\meter} PMMA spheres, for granule volume fractions ranging from $\phi_\text{G} =20~\%$ to $\phi_\text{G} =45~\%$. Main graphs: thickening index as a function of the shear stress. Insets: Viscosity as a function of shear rate.
    Black curves represent the baseline behavior of the pure fumed silica suspension extracted from Fig.\ref{fig1} at $\phi_\text{STF} = 30\,\%$.}
    \label{fig4}
\end{figure}
The effects of granule size, polydispersity, and surface properties are now investigated using larger PMMA spheres and smaller, polydisperse glass spheres.
In the same fumed silica suspension as used in Fig.\ref{fig3}, monodisperse 60-\si{\micro\meter} and polydisperse 10-\si{\micro\meter} spheres are combined, and their rheology are shown in Fig.\ref{fig4}$(a,\,b)$, respectively.
The volume fraction of the granules, $\phi_\text{G}$, is varied over the same range as in Fig.~\ref{fig3}. 
Similar trends are observed in both cases, as shown by the thickening index plotted against shear stress (main graphs) and the viscosity curves as a function of shear rate (insets).
Again, the shear rate at the onset of shear thickening decreases with increasing granule volume fraction, consistent with earlier findings.
Also, the shear stress at the onset of shear thickening, $\sigma_\text{c}$ (defined as the stress where $\beta = 0$), remains nearly constant regardless of the granule volume fraction, as evidenced in Fig.~\ref{fig4}.
These results further confirm the generality of the trends observed with different granule types and sizes.

\begin{figure}
    \centering
    \includegraphics[width=0.8\linewidth]{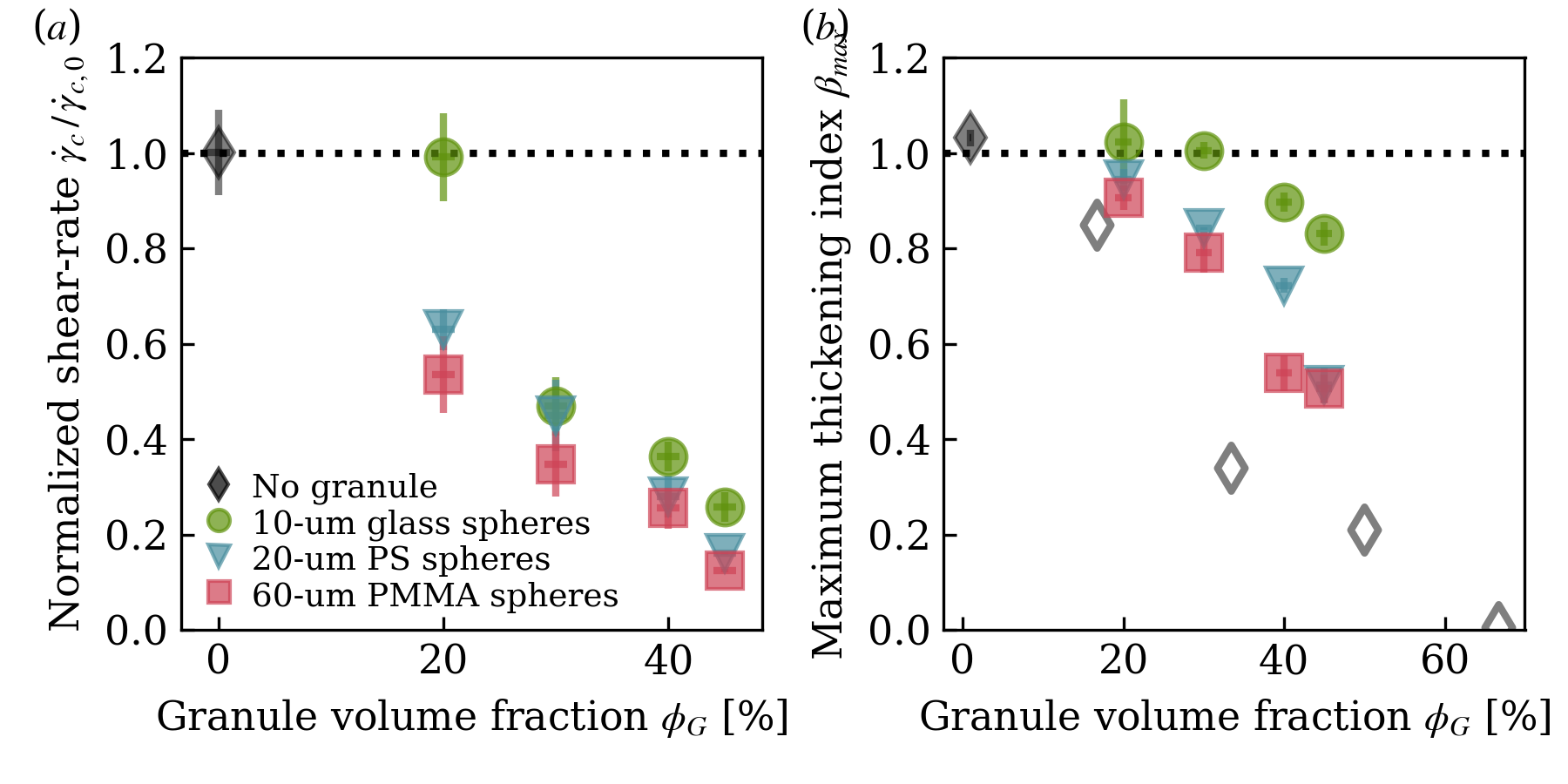}
    \caption{Summary of the shear thickening features of a mixture of a DST suspension of fumed silica $\phi_\text{STF} = 30\%$, with much larger granules (see legend). ($a$) Shear rate at the onset of thickening behavior and ($b$) Maximum thickening index. 
    Open gray lozenges are measurements in pure fumed silica suspensions from Fig.~\ref{fig1}(c).
    }
    \label{fig5}
\end{figure}

The shear-thickening characteristics for different granule  sizes and granule volume fractions in the same fumed silica suspension are summarized in Fig.~\ref{fig5}.
In Fig.~\ref{fig5}(a), the shear rate at the onset of shear thickening, $\dot{\gamma}_\text{c}$, is plotted as a function of the granule volume fraction, $\phi_\text{G}$. 
The data confirm that higher granule volume fractions result in earlier shear-thickening onset. 
Additionally, this figure highlights the influence of granule size: larger granules lead to a greater shift in the effective shear rate.
These differences between fumed silica-based and cornstarch-based systems are explored further in the discussion section.

Fig.~\ref{fig5}(b) presents the maximum thickening index, $\beta_\text{max}$, for the various suspensions. 
The trend confirms earlier observations: shear thickening becomes weaker as either the granule size or volume fraction increases. 

\begin{figure}
    \centering
    \includegraphics[width=0.5\linewidth]{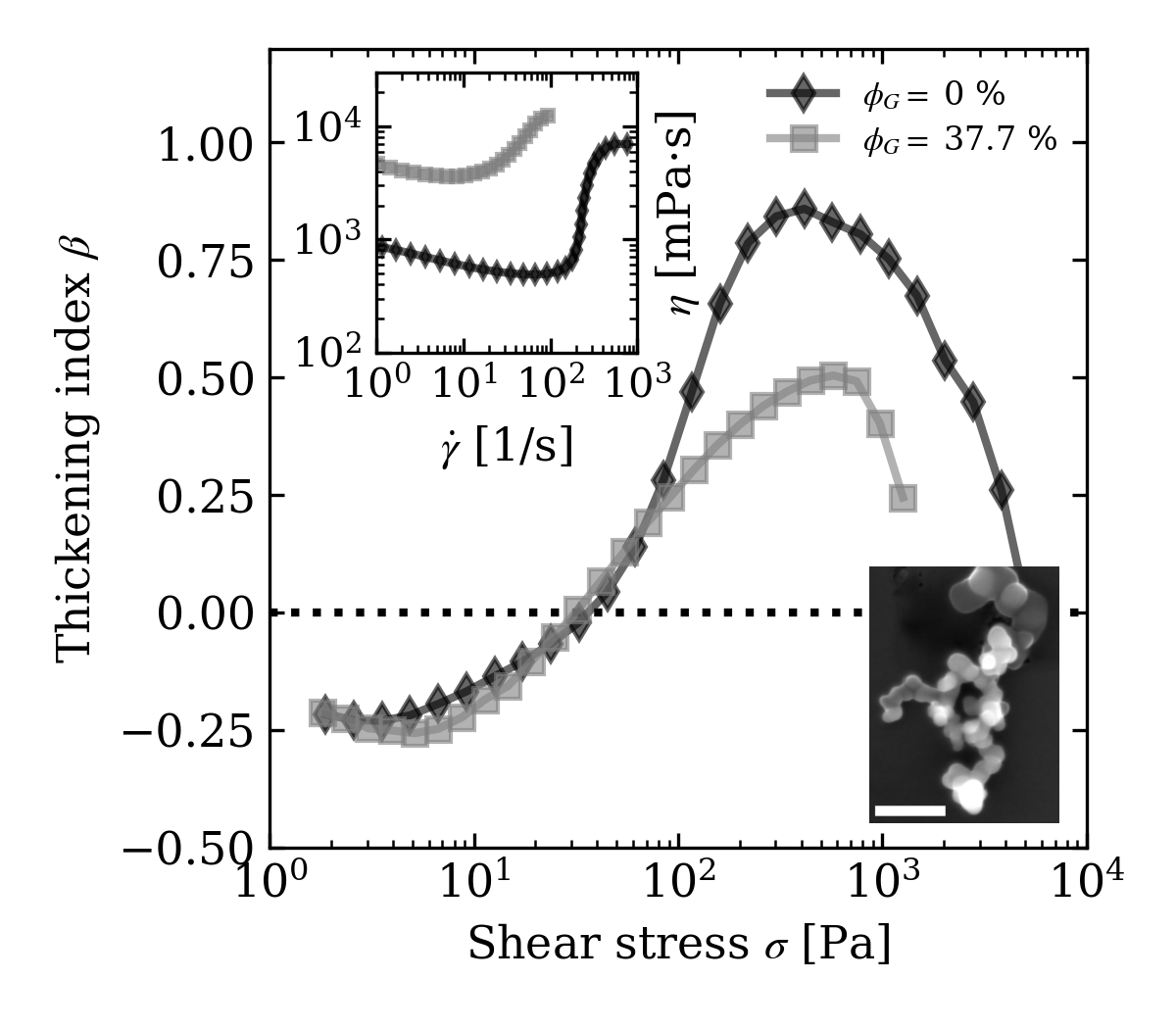}
    \caption{Rheology of suspension made with fumed silica (bottom right SEM picture, scale: \SI{200}{\nano\meter}) in PEG200, $\phi_\text{STF} =25.5~\%$ combined with 60-\si{\micro\meter} PMMA spheres, for granule volume fractions $\phi_\text{G} =37.7~\%$. Main: thickening index as a function of shear-stress. Inset: steady shear viscosity as a function of shear-rate. }
    \label{fig:FS25}
\end{figure}

\textbf{CST fluid and granules -}
The same trends are observed for a continuously shear-thickening (CST) suspension of fumed silica at lower volume fraction.
At $\phi_\text{STF}=25.5\,\%$, the thickening index with no granule is $\beta_\text{max}$ = 0.85, and drops to $\beta_\text{max} = 0.5$ when 60-\si{\micro\meter} are added at $\phi_\text{G} = 30\,\%$, see Figure~\ref{fig:FS25}.

\begin{figure}
    \centering
    \includegraphics[width=0.5\linewidth]{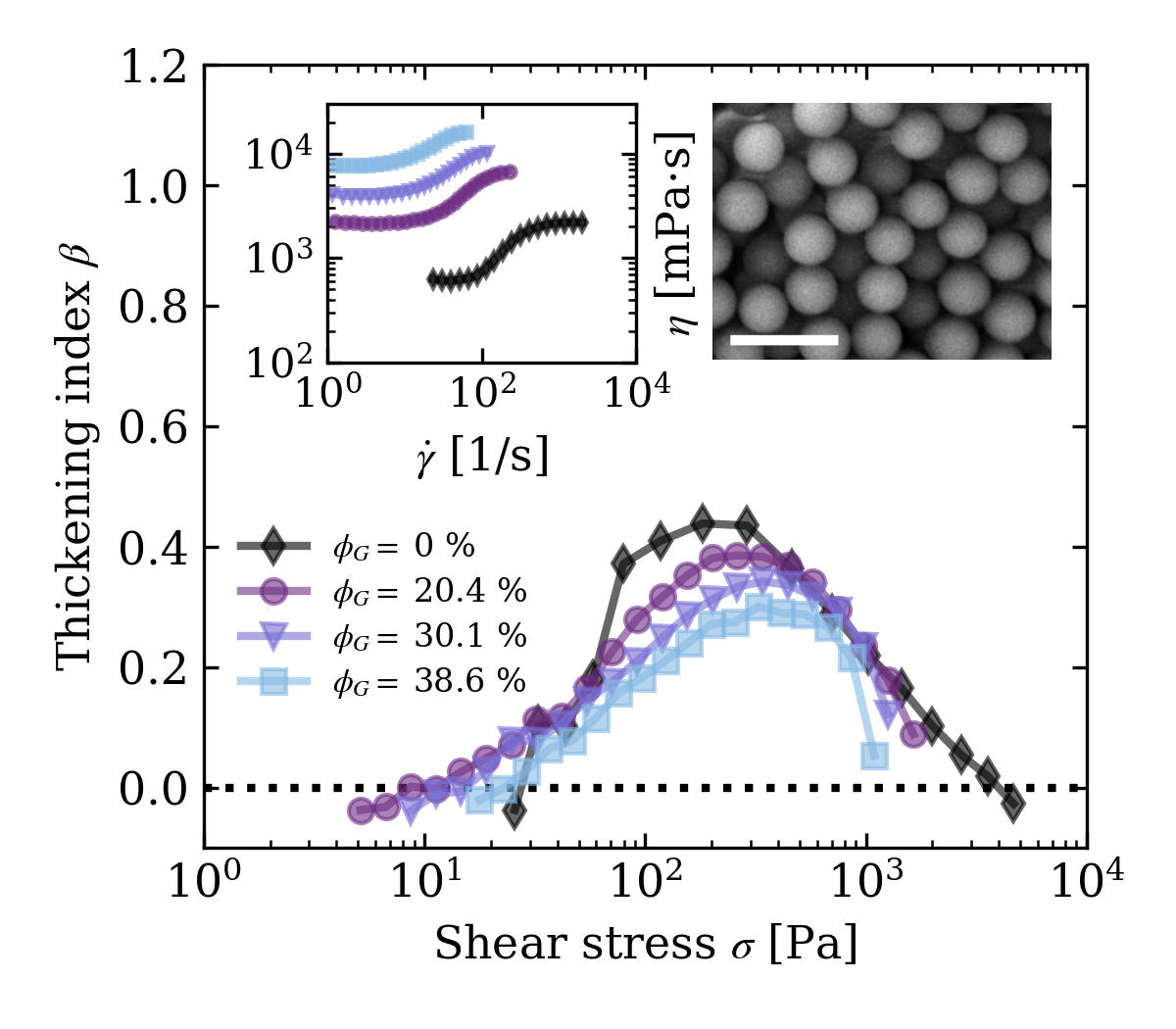}
    \caption{Rheology of suspension made with a colloidal silica spheres (upper right SEM picture, scale: \SI{1}{\micro\meter}) in PEG200, $\phi_\text{STF} =52~\%$ combined with 60-\si{\micro\meter} PMMA spheres, for granule volume fractions ranging from $\phi_\text{G} =20~\%$ to $\phi_\text{G} =40~\%$ ($a$) Steady shear viscosity as a function of shear-rate. ($b$)Thickening index as a function of shear-stress.}
    \label{fig:nano-spheres}
\end{figure}

We also investigated a shear-thickening fluid made of 500-\si{\nano\meter} silica spheres in PEG200 with $\phi_\text{STF} = 52\%$ to test the generality of the results found with fumed silica.
Similarly to the low concentration fumed silica suspension, the baseline viscosity exhibits CST as shown in Fig.\ref{fig:nano-spheres}.
The maximum thickening index of this suspension is $\beta = 0.5$.
The rheology upon addition of 60-\si{\micro\meter} PMMA granules in this suspension is presented in Fig.\ref{fig:nano-spheres} for different loading of granules.
Again, the thickening index decreases upon addition of granules.

These experiments with CST suspensions made of fumed silica and silica nanospheres thus show trends similar to those found in DST fumed silica suspension, namely (i) an increase of the zero-shear viscosity, (ii) an onset of shear-thickening at smaller shear rates, and (iii) a milder shear-thickening.
The latter is seen by the decrease in the maximum value of the thickening index in the insets of Fig.~\ref{fig:FS25} and Fig.~\ref{fig:nano-spheres}.
Similar results are seen at higher volume fractions of silica nanospheres.


\section{Discussion, Tentative Scenario and Conclusions}
\label{sec:discussion}

To summarize our results, for shear-thickening fluids made of different nanoparticles, we find that adding  20 to 120 times larger spherical granules shifts the shear-thickening onset toward smaller shear rates while it does not affect the onset shear stress.
Importantly, we observe that the shear-thickening becomes softer when the large granules are added and quantified this using $\beta_\text{max}$.
This softening is seen for silica nanoparticle suspensions of different particle shape (very rough, non-spherical fumed silica particles as well as smooth silica spheres) and different initial degrees of shear-thickening in the absence of the granules (both CST and DST) as well as for different granule sizes and materials.
Despite the large size contrast between granules and nanoparticles, granule size remains important in determining the degree of this softening. 
We now discuss and compare these results with those found in the existing literature.

In general, for suspensions with a bimodal particle size distribution of small and large particles we can expect that the rheology transitions from the behavior of the small to that of the large particles alone, as the proportion of small particles in the solid phase, $\alpha$, decreases at fixed total solid volume fraction \citep{bender1996reversible, singh2024rheology}. 
As far as the non-shear-thickening regimes of the flow curves are concerned, the role of the granules then merely is to shift the suspension viscosity upward \citep{sengun1989bimodal}.
In the shear-thickening regime, the onset stress scales inversely with the particle size $d$ according to $\sigma_\text{c} \propto d^{-\gamma}$ with exponent $\gamma$ in the range 1.75 to 3 \citep{bender1996reversible, maranzano2001effects, guy2015towards,kim2024shear}.
We thus expect $\sigma_\text{c}$ to shift to lower values as soon as the large particles contribute significantly to the thickening, provided they do. 
In our experiments the fraction of granules was varied from $\phi_\text{G} = 0\,\%$ up to $\phi_\text{G} = 45\,\%$, in which case nanoparticle represent only a quarter of the total solid volume ($\alpha = 27\,\%$, see Table~\ref{tab1}).
The finding that $\sigma_\text{c}$ is not varying with the relative proportion of small to large particles for $\alpha$ ranging from 100~\% to 27~\% therefore indicates that only the small nanoparticles trigger shear-thickening.
That we do not observe a noticeable lowering of $\sigma_\text{c}$ even for $\alpha$ = 27~\% differs from recent simulations of binary non-Brownian suspensions \citep{singh2024rheology}, where the onset stress decreases from that of a suspension made of only small particles once $\alpha\leq 0.5$. 
To conclude, this fixed value of $\sigma_\text{c}$ varying $d_\text{G}$ and $\phi_\text{G}$ strongly suggests that granule contribution to the shear-thickening process is negligible in comparison to that of the nanoparticles. 

The softening of the shear thickening we observe in silica nanoparticle suspensions when granules are added differs significantly from that reported in the literature for cornstarch suspensions \citep{madraki2017enhancing,madraki2018transition}: there, granule addition enhances the shear-thickening behavior and drives a transition from CST to DST.
\cite{madraki2018transition} found that their observed enhancement of shear thickening could be explained by an effective increase in the volume fraction of the small particles from $\phi_\text{STF}$ to $\phi_\text{STF,eff}$ due to the presence of the granules.  
When the size ratio between granules and shear-thickening small particles is not very large, excluded volume near the surface of the granules prevents small particles from packing at their bulk density.  This geometric `wall effect’ enhances the effective concentration in the interstitial fluid according to $\phi_\text{STF, eff}/\phi_\text{STF}= V_\text{total}/(V_\text{total}-V_\text{shell})\sim 1+V_\text{shell}/V_\text{total}$, where $V_\text{shell}$ is the excluded volume of the shell around the granules that depends as $V_\text{shell}/V_\text{total} = \phi_\text{G}\left[\left(1+d_\text{STF}/d_\text{G}\right)^3-1\right]\sim3\phi_\text{G} d_\text{STF}/d_\text{G}$ on the diameters of the shear-thickening particles, $d_\text{STF}$, and the granules, $d_\text{G}$.
In the experiments by \cite{madraki2018transition} the granules were no larger than 10 times the size of the cornstarch particles (typically \SI{20}{\micro\meter}), so that an appreciable enhancement $\phi_\text{STF, eff}/\phi_\text{STF} > 1$ due to the wall effect could be observed.    
However, in our experiments the characteristic size of the shear-thickening fumed silica particles  is \SI{500}{\nano\meter} and the smallest granules are \SI{10}{\micro\meter} such that $d_\text{STF}/d_\text{G}\leq 0.05$, $V_\text{shell}/V_\text{total}\ll  1$ and $\phi_\text{STF, eff}/\phi_\text{STF}\simeq 1$. 
Therefore, the wall effect cannot affect the shear thickening propensity in our experiments.

A weakening of the shear-thickening in highly bimodal suspensions was found in simulations of non-Brownian particles when the total solid volume fraction was kept fixed and small particles were replaced by an equal volume of large ones \citep{malbranche2023shear,singh2024rheology}.
For small size ratios (up to 10) this is explained by an increased distance to the jamming volume fraction with bimodal systems.
For the large size ratio in our systems, the dependence of the jamming volume fraction on the size ratio saturates, and the variations observed in Fig.\ref{fig5}(b) for instance, cannot be explained by an increase in the distance to the jamming volume fraction.

One key question thus remains: how can the granule size leave such a clear signature in the rheological behavior of the suspension considering that any granule is much larger than the nanoparticles?
In particular, why does the strength of shear thickening decrease, and why is the decrease more pronounced as the granule size increases?
We identify two aspects that could account for these observations.
First, as mentioned earlier, the addition of granules increases the low-shear viscosity by a factor of $\eta_r(\phi_\text{G})$, which is independent of the granule size. 
Additionally, and more importantly for shear thickening, it leads to a local enhancement of the shear rate experienced by the nanoparticles within the interstices among the granules.
On average, this local enhancement is quantified by the lever function, $\mathcal{F}(\phi_\text{G})$, such that $\dot{\gamma}_\text{local}/\dot{\gamma}_\text{macro} = \mathcal{F}(\phi_\text{G})$.
Notably, this average enhancement is independent of granule size.
Unlike a pure nanoparticle suspension sheared uniformly by the walls of a rheometer, this local shear rate within the suspension is non-homogeneous and fluctuates due to granule motion. 
This heterogeneity in the shear rate could partially explain the observed softening, as it gives rise to localized thickening events across the suspension at a given macroscopic shear rate. 
However, the behavior of local shear-rate fluctuations is expected to mirror that of the mean local shear rate, which is independent of granule size.
This does not capture the granule size dependent trends we see in Fig.\ref{fig5}($b$)
We thus propose a tentative scenario based on comparing the size of the largest frictionally interacting nanoparticle cluster, $d_\text{cluster}$, to that of the granules $d_\text{G}$ to account for the granule size effect.
In this scenario, clusters grow undisturbed within the granule interstices when $d_\text{cluster}<d_\text{G}$, envelop granules if $d_\text{cluster}>d_\text{G}$, but interact with and are disrupted by granules of comparable size. 
For discontinuous shear thickening (DST) to occur, a cluster must span the entire system. 
The cluster size depends on the applied shear rate and the friction coefficient.
While small clusters can form spontaneously in many regions of the suspension, the growth of larger clusters becomes increasingly constrained. 
Disruption of large clusters by large granules is particularly detrimental to the formation of force chains, thereby reducing the intensity of shear thickening. 
This explains the observed decrease in $\beta_\text{max}$ with increasing granule size, as shown in Fig.\ref{fig5}($a$).
In other systems, the disruption of cluster growth, i.e., the break-up of force chains, by ultrasound \citep{sehgal2019using} or orthogonal oscillations \citep{lin2016tunable} for instance, has been shown to inhibit the shear-thickening process.
We hypothesize that, in the present case, the interaction and subsequent disruption of growing clusters by large granules are responsible for the observed size-dependent softening of shear thickening.\\

\noindent\textbf{Safety and hazards}\\
Dry fumed silica is easily aerosolized and extremely hazardous
to the respiratory system, and must be handled with precautions inside of a fume hood. See OSHA guidelines for detailed regulations.\\

\noindent\textbf{Conflicts of interest}\\
There are no conflicts to declare.\\

\noindent\textbf{Acknowledgments} \\
Data, codes and scripts necessary to reproduce the results reported in this article are available on request. 
AP and HMJ acknowledge support from the
Army Research Office under grant W911NF-21-2-0146.
AP acknowledges the University of Chicago Materials Research Science and Engineering Center, funded
by the National Science Foundation under award number DMR-
2011854, for the Kadanoff-Rice Postdoctoral Fellowship.  

\bibliographystyle{apalike}

\bibliography{biblio-STF-granules}

\end{document}